\documentclass[twocolumn,10pt]{IEEEtran}

\usepackage{bm,cite,float,amsmath,amssymb,amsthm}
\usepackage{bytefield}
\usepackage{algorithm}
\usepackage[noend]{algpseudocode}
\usepackage{indentfirst}
\usepackage{xcolor}

\makeatletter
\def\BState{\State\hskip-\ALG@thistlm}
\makeatother

\usepackage{amssymb}
\usepackage{amsmath}
\usepackage{graphicx}
\usepackage{cite}

\usepackage{balance}
\usepackage[utf8]{inputenc}

\IEEEoverridecommandlockouts

\usepackage{graphicx,epstopdf}
\usepackage{epsfig}	
\usepackage{amsfonts,balance}
\usepackage{bbm}
\floatname{algorithm}{Algorithm}
\usepackage{lipsum}

\newtheorem{theorem}{Theorem}

\newtheorem{corollary}{Corollary}

\makeatletter
\def\ScaleIfNeeded{%
	\ifdim\Gin@nat@width>\linewidth \linewidth \else \Gin@nat@width
	\fi } \makeatother

\begin{document}
	
	\title{Real-time Video Streaming and Control of Cellular-Connected UAV System: \\Prototype and Performance Evaluation
	}
	
	\author{Hui~Zhou,~\IEEEmembership{Student Member,~IEEE,}
		Fenghe Hu,~\IEEEmembership{Student Member,~IEEE,}
		 Michal~Juras,\\
		Asish B Mehta,
		and Yansha Deng,~\IEEEmembership{Member,~IEEE}
		
		\thanks{Hui~Zhou, Fenghe Hu, Michal~Juras, Asish B Mehta and Yansha Deng are with Department of Engineering, King's College London, London, WC2R 2LS, UK (email:\{hui.zhou, fenghe.hu, michal.juras\}@kcl.ac.uk, asishm.school@gmail.com, yansha.deng@kcl.ac.uk)(Corresponding author: Yansha Deng).}
		
	}
	
	\maketitle
	
	\begin{abstract}
		Unmanned aerial vehicles (UAVs) play an increasingly important role in military, public, and civilian applications, where providing connectivity to UAVs is crucial for its real-time control, video streaming, and data collection. Considering that cellular networks offer wide area, high speed, and secure wireless connectivity, cellular-connected UAVs have been considered as an appealing solution to provide UAV connectivity with enhanced reliability, coverage, throughput, and security. Due to the nature of UAVs mobility, the throughput, reliability and End-to-End (E2E) delay of UAVs communication under various flight heights, video resolutions, and transmission frequencies remain unknown. To evaluate these parameters, we develop a cellular-connected UAV testbed based on the Long Term Evolution (LTE) network with its uplink video transmission and downlink control\&command (CC) transmission. We also design algorithms for sending control signal and controlling UAV. The indoor experimental results provide fundamental insights for the cellular-connected UAV system design from the perspective of transmission frequency, adaptability, and link outage, respectively.
		
	\end{abstract}
	
	
	\section{Introduction}
	Unmanned Aerial Vehicles (UAVs) with low cost and high mobility are an emerging technology and have found a wide range of applications, including food delivery, public safety, traffic monitoring, and many others over the past few decades\cite{Zeng2016,Challita2019a,Yang2019,Fan2019}. As these applications require reliable control of UAVs and real-time application data transmission, there is an urgent need for wireless technology that can guarantee ultra-reliable low latency communication (URLLC) for downlink command\&control (CC) transmission, and high throughput low E2E delay for mission-related data transmission \cite{Azari2019}.
	
	Conventional UAVs communication simply relies on the short-range communications (e.g., WiFi, and Bluetooth) with short transmission range, inefficient multi-UAVs collaboration, and limited multi-UAVs control. This may not be sufficient for beyond visual line-of-sight (LOS) communication needs, particularly for those applications requiring wide-area connectivity. To tackle this, cellular-connected UAVs communication has been proposed to allow beyond LOS control, low E2E delay, real-time communication, robust security, and ubiquitous coverage to support a myriad of applications ranging from real-time video streaming to surveillance.
	
	Most existing work mainly focused on the analysis and simulation of cellular-connected UAVs \cite{lin2018,Azari2019,Zeng2019}. In \cite{Azari2019}, a comprehensive analysis of coverage and data rate was carried out for the downlink. In \cite{lin2018}, simulation results of uplink throughput were obtained to exploit the feasibility of providing LTE connectivity for UAVs. In \cite{Zeng2019}, downlink data rates with fixed antenna pattern and 3D beamforming were compared. However, theoretical papers mainly focused on the physical layer without considering the content of transmission and protocols in cross-layers. Moreover, the mathematical channel model can not capture the full characteristics of practical implementation such as antenna angle and pattern, packet transmission frequency, obstacles, and etc.
		
	To evaluate the E2E performance of downlink control and uplink video transmission, we develop a cellular-connected UAV testbed based on the LTE network, which consists of the physical layer, MAC layer, network layer, transport layer, and application layer, and can be easily extended to 5G. The LTE network is established via OpenAirInterface (OAI) \cite{OAI} and Software Defined Radio (SDR). Then, we develop the downlink CC transmission algorithms and real-time uplink video streaming transmission through the LTE network. We also propose methods to measure the E2E delay of the CC transmission and video transmission. Finally, we carry out indoor experiments and analyze the experimental results for various flight heights, video resolutions, and transmission frequencies. The contributions of this paper are twofold. First, to the best of the authors' knowledge, it is the first paper that built the cellular-connected UAV testbed for real-time video streaming and control signal transmission. Second, the experimental results provide insights for the design of cellular-connected UAV system in practice as follows:
	\begin{itemize}
		\item Although higher transmission frequency offers more precise and smooth control of the UAV-UE, higher transmission frequency than a certain threshold leads to buffer overflow and significant increase of latency, which cannot guarantee the operation safety;
		
		\item Due to the mobility of the UAV-UE, the UAV network parameters, including locations, channel status, and etc, easily change during the missions. Therefore, it is crucial for the UAV system to dynamically adjust FPS to maintain continuity of the video transmission;
		
		\item As the antennas in the BS are tilted downward to the ground users, this provides limited gain for the UAV-UE and results in lower throughput, especially when UAV-UE passing through the top of the BS. Therefore, link outage in the cellular-connected UAV system remains an open problem to be solved;
		
	\end{itemize}
	Note that all the above insights can only be obtained from performing our experimental test, rather than theoretical analysis.

	The rest of this work is organized as follows. Section II presents the system overview and performance metrics. Section III provides the testbed setup. Section IV presents the experimental results. Finally, Section V concludes the paper.
	
	\section{System Overview}
	
	We consider cellular-connected UAVs to support various real-time video streaming applications, such as inspection, and product delivery, where uplink and downlink are established between the UAV User Equipment (UAV-UE) and the BS for video streaming and CC transmission, separately \cite{BorYaliniz2019}. Knowing that ultra-reliable and low latency are main performance metrics to guarantee the UAV safe control, high throughput and low latency are important for real-time video streaming, we formulate these metrics from the perspective of uplink video transmission and downlink CC transmission as following:
	

	\subsubsection{E2E downlink CC delay}
	The E2E CC link delay $ D_{\mathrm{CC}} $ is defined from the Evolved Packet Core (EPC) to the UAV-UE, which is the sum of transmission delay, propagation delay, processing delay, and queueing delay. To do so, we perform clock synchronization between the EPC and the UAV-UE to eliminate clock error, which will be explained in section III in detail. The E2E delay of downlink CC transmission $ D_{\mathrm{CC}} $ can be expressed as
	\begin{equation}
	D_{\mathrm{CC}} = T_{\mathrm{Cr}} - T_{\mathrm{Ct}},
	\label{CC link latency}
	\end{equation}
	where $ T_{\mathrm{Ct}} $ is the transmit timestamp of the CC recorded at the EPC, $ T_{\mathrm{Cr}} $ is the received timestamp of the CC recorded at the UAV-UE.
	
	\subsubsection{E2E uplink video delay} The E2E delay of video transmission $ D_{\mathrm{Vd}} $ represents the amount of time it takes for a single frame of video to transmit from the camera at the UAV-UE to the display at the EPC, which is formulated as:
	\begin{equation}
	D_{\mathrm{Vd}} = T_{\mathrm{Vr}} - T_{\mathrm{Vt}},
	\label{Application link latency}
	\end{equation}
	where $ T_{\mathrm{Vt}} $ is the time when the typical video frame is captured by the camera, $ T_{\mathrm{Vr}} $ is the time when the typical video frame is shown on the screen at the EPC.

	\subsubsection{Downlink CC transmission reliability}
	The downlink CC transmission reliability $ R_{\mathrm{CC}} $ is calculated as
	\begin{equation}
	R_{\mathrm{CC}} = \dfrac{N^{\mathrm{CC}}_{\mathrm{rece}}}{N^{\mathrm{CC}}_{\mathrm{trans}}},
	\label{CC link data rate}
	\end{equation}
	where $ N^{\mathrm{CC}}_{\mathrm{rece}} $ and $N^{\mathrm{CC}}_{\mathrm{trans}}$ are the number of successfully received CC and total transmitted CC, respectively.
	
	\subsubsection{Uplink video transmission throughput}
	The video transmission throughput $ T_{\mathrm{Vd}} $ can be calculated based on the video frame E2E delay and the video frame size, where the video frame size $ S_{\mathrm{Vd}} $ varies for different video resolutions and video encoding formats. Thus, the video transmission throughput $ T_{\mathrm{Vd}} $ can be evaluated using
	\begin{equation}
	T_{\mathrm{Vd}} = \dfrac{S_{\mathrm{Vd}}}{D_{\mathrm{Vd}}}.
	\label{Application link data rate}
	\end{equation}

	\section{Testbed Setup}
	To evaluate the reliability, throughput and E2E delay, we build a cellular-connected UAV testbed as described in this section. In detail, we first describe the hardware part including the UAV-UE and ground control station (GCS) setup, then describe the software configuration implemented based on OAI, and at last present the communication part including the downlink CC transmission and the uplink video transmission.
	
	\subsection{Hardware}\label{AA}
	\subsubsection{\textbf{UAV-UE Setup}}
	As shown in Fig.~\ref{Hardware Setup} (a), The DJI MATRICE 100 \cite{dji2020} is selected as the UAV-UE, as it supports additional expansion bays to customize the payload and universal communication ports for connecting third-party components. The payload is fixed on top of the UAV-UE, which consists of the following parts:
	
	
	\begin{figure}[!tb]
		\centerline{\includegraphics[scale=0.28]{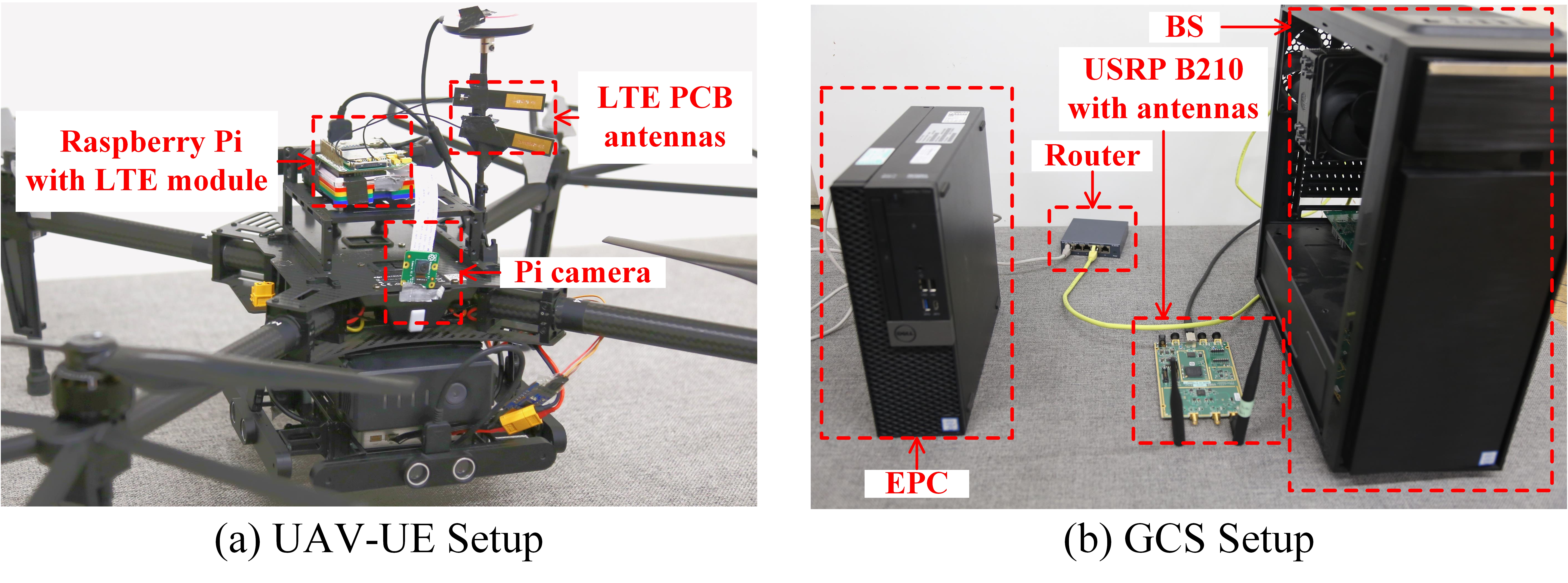}}
		\caption{Hardware setup.}
		\label{Hardware Setup}
	\end{figure}
	
	\begin{itemize}
		\item Raspberry Pi: Raspberry Pi 4 is chosen as the UAV-UE onboard computer because of its lightweight. It receives the CC frame from the BS wirelessly and routes the CC to the UAV-UE through USB to UART adapter. It also encodes the video and transmits the real-time video streaming to the BS.
		\item LTE Module: To facilitate the UAV-UE communication with the BS, the Quectel EC25 LTE module is installed on the Pi through an interface bridge for LTE communication. Two LTE PCB antennas are installed on the LTE module to enable Maximum Ratio Combination (MRC) at the UAV-UE.
		
		
		
		\item Camera: The Raspberry Pi camera module is selected for video capturing due to its compatibility with Pi and the capability of capturing high-definition video. 
		
	\end{itemize}

	\subsubsection{\textbf{GCS Setup}}
	The GCS consists of the remote controller, EPC, and the BS, as shown in Fig.~\ref{Hardware Setup} (b). The third-party remote controller is selected to be wired connected with the EPC to use the established LTE network for UAV-UE communication instead of DJI's embedded WiFi module. The EPC and the BS are installed on two different computers connected via Ethernet and the BS is equipped with Universal Software Radio Peripheral (USRP) as a radio frequency (RF) unit.
	
	\begin{itemize}
		\item USRP: USRP B210 \cite{ETTUS2020} is selected for radio transmission and reception in our testbed. It can provide a fully integrated, single-board platform with continuous frequency coverage from 70 MHz – 6 GHz. 
		
		\item Antenna: Two omnidirectional antennas with 3dBi gain are installed on the TX/RX ports of USRP B210, which support dual-band.
		\item PCs: The EPC is set up on the PC with i7-7700 CPU and 16GB memory, and the BS is set up on the other PC with i9-9900K CPU and 48GB memory due to the heavy use of integer Single Instruction Multiple Data (SIMD) instructions.
		
		
	\end{itemize}
	
	\subsection{Software}
	We select an opensource project OAI, which consists of software EPC, BS, and UE, to implement the LTE network. In a general EPC, Home Subscriber Server (HSS) holds the database, which contains information related to UE authentication and access authorization. The Mobility Management Entity (MME) mainly controls the mobility and security for access. The Serving Gateway (S-GW), and Packet Data Network Gateway (P-GW) mainly act as interfaces that can serve UEs by routing the incoming and outcoming IP packets. A programmable SIM card is selected in the LTE module to match the database registered in the HSS of the EPC. After correctly configuring the parameters in the SIM card, such as security key, registration information, and corresponding gateway, the UAV-UE can attach to the configured BS and then connect with the EPC.

\begin{figure}[!htb]
	\centerline{\includegraphics[scale=0.6]{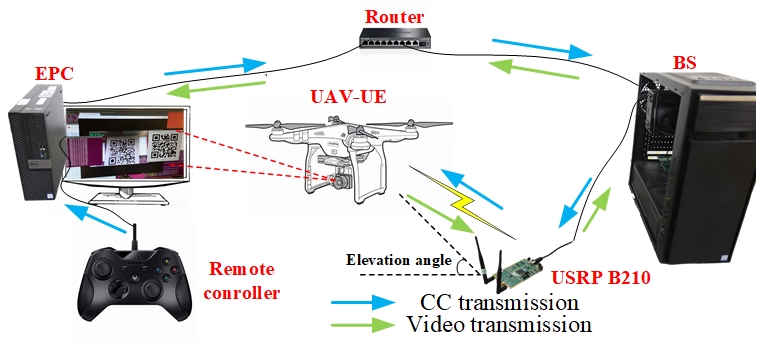}}
	\caption{Diagram of cellular-connected UAV testbed.}
	\label{Diagram of Cellular-connected UAV Testbed}
\end{figure}
	
	\subsection{Communication}
	Our cellular-connected UAV communication testbed is shown in Fig.~\ref{Diagram of Cellular-connected UAV Testbed}, and we will present the downlink CC transmission and uplink video transmission in this part. Before the establishment of the links, we synchronize the time between the EPC and the UAV-UE for the E2E delay measurement. The time synchronization is achieved through Network Time Protocol (NTP) and the NTP server query interval is 10 seconds, which can guarantee the time synchronization error less than 1ms.

	\subsubsection{\textbf{Downlink CC transmission}}
	
	The downlink CC transmission consists of two processes: sending control signal, and controlling UAV-UE. First, the EPC encodes the CC signal from the remote controller and sends it to the UAV-UE using the established LTE network. After that, the UAV-UE decodes the CC signal and controls the UAV-UE movement through the DJI API. We present the algorithms for sending control signal in Algorithm 1, and controlling UAV-UE in Algorithm 2. In the following, we describe the details of these two algorithms.
	\begin{algorithm}
		\caption{The procedure of sending control signal}
		\label{control-sender}
		\begin{algorithmic}[1]
			\Procedure{control signal sending}{}
			\State Initialise UDP socket
			\State Initialise Joystick
			\State Setting destination IPV4 address
			\Loop
			\State Obtain CC parameters $roll,pitch,yaw,thrust$
			\State Normalize CC parameters value
			\State Encode normalized CC parameters
			\State Send CC frame to destination
			\State Record frame ID and transmit time
			\State Wait for pre-set time
			\EndLoop
			\State Close socket
			\EndProcedure
		\end{algorithmic} 
	\end{algorithm} 
	
	The Algorithm 1 begins by creating a UDP socket on the EPC to send data through. There are two reasons to select non-blocking UDP protocol in the cellular-connected UAV testbed. One is that the control signal has to be sent at exact intervals without retransmission, and the other reason is to reduce the delay introduced by handshake when operating with stateful protocols like TCP. After that, we initialize the remote controller, which is a wired Xbox joystick in the case of our implementation. 
	

	After initialization, we keep monitoring the CC from the remote controller, and four UAV-UE movement parameters are defined to control the UAV-UE. Roll, pitch, and yaw control the UAV-UE to rotate in three dimensions, and thrust controls the UAV-UE to move up or down. The CC parameters are normalized to map inside the range of UAV-UE parameters for smooth control, and are then encoded into the predefined frame structure, which includes four movement parameters along with the frame ID, and each of them occupies 4 Bytes. Finally, the CC frame is sent at a particular frequency, which is lower than 50Hz to prevent buffer overflow on the UAV-UE side. Once a CC signal has been sent, the EPC will record the corresponding frame ID and transmit time into the local file for E2E delay evaluation.
	

	\begin{algorithm}
		\caption{The procedure of controlling UAV-UE}
		\label{control-receiver}
		\begin{algorithmic}[1]
			\Procedure{Controlling UAV-UE}{}
			\State Initialize UDP socket
			\State Initialize UAV control API
			\State Set the control flag of the UAV
			\Loop
			\State Receive the CC frame $ data $
			\If {$data$ is not $NULL$} 
			\State Set DJI API parameters to $ data $
			\Else
			\State Set DJI API to zeros
			\EndIf
			\State Send parameters and control flag through API
			\State Record frame ID and received time
			\State Sleep for 20 ms
			\EndLoop
			\State Close socket
			\EndProcedure
		\end{algorithmic}
	\end{algorithm}
	
	The Algorithm 2 running at the UAV-UE starts with the initialization of the UDP socket and DJI control API. We also set the control flag to indicate the intention to control the UAV-UE with the third-party remote controller. After that, the UAV-UE keeps monitoring the traffic from the UDP port and parses the CC frame to obtain the roll, pitch, yaw, thrust, and the frame ID. If the obtained data is not null, the input parameters of DJI API is set to the obtained data, otherwise, it is set to zeros. Finally, the corresponding frame ID and received time are recorded at the UAV-UE. As suggested by the DJI, the program remains idle for 20 ms and then parses the received CC frame again to achieve 50Hz control frequency.
	
	\subsubsection{\textbf{Uplink video transmission}}
	To allow the real-time video streaming transmission, an open-source server WebRTC\cite{WebRTC} is installed on the Raspberry Pi, and the EPC fetches real-time video streaming from 8080 port of the Raspberry Pi. WebRTC adopts the codec based on H.264, and allows to adjust FPS automatically.
	
	To measure the E2E delay, we apply the QR-code to carry the transmit and the receive time of video frame\cite{Boyaci2009}. The E2E delay measurement scheme consists of two processes:  QR-code generation and QR-code recognition. During the QR-code generation process, the QR-code generation program runs at the EPC and encodes the local time to QR-code with a 60Hz refresh rate. The generated QR-code is shown on the right side of the screen at the EPC and captured by the UAV-UE's camera. During the QR-code recognition process, the EPC receives the video frame representing the transmit time from the UAV-UE, and displays the QR-code at the left side of the screen. The other program running at the EPC takes a screenshot with a frequency of 60Hz, and recognizes the QR-code on both sides to obtain the transmit time and the receive time. Finally, the E2E delay can be calculated based on \eqref{Application link latency}.

	\section{Experiments}
	In this section, we carry out the experiments to evaluate the throughput, reliability and E2E delay of uplink video transmission and downlink CC transmission under various UAV-UE flight heights, CC transmission frequencies and video resolutions. Due to the strict restrictions of Federal Aviation Administration (FAA), we conduct indoor experiments for three UAV-UE flight height $ h=$0m, 1m, 2m, where $h=0m$ represents the stationary UAV-UE on the ground. However, with the algorithms and practical implementations performed for this real-time cellular-connected UAV testbed, the experiment can be easily extended to multiple UAV scenarios and higher heights with the regulation approval for outdoor flight test. In all figures of this section, we use “Ave” to
	abbreviate “Average”. 
	\begin{table}[!htb]
		\caption{Bandwidth and round-trip delay LTE network}
		\begin{center}
			\begin{tabular}{|c|c|c|}
				\hline
				& 25 PRB & 50 PRB \\
				\hline
				Uplink bandwidth (Mb/s)& 8.78 & 18.77 \\
				\hline
				Downlink bandwidth (Mb/s)& 16.57 & 34.3 \\
				\hline
				Round-trip delay (ms)& 27.66 & 29 \\
				\hline
			\end{tabular}
			\label{Performance of the the established LTE network}
		\end{center}
	\end{table}

	We first use iperf \cite{iperf} and ping to measure the bandwidth and round-trip delay of the established LTE network, and the detailed results under 25 physical resource block (PRB) and 50 PRB are given in Table~\ref{Performance of the the established LTE network}. Both uplink and downlink bandwidth with 50 PRB are more than twice the bandwidth of 25 PRB, and the round-trip delay is almost the same. Although 50 PRB can support higher bandwidth, its uplink and downlink are not stable compared to the LTE network with 25 PRB in OAI. Therefore, an LTE network with 25 PRB is established for the following CC and video transmission experiments. 
	
	\begin{figure}[!tb]
		\centerline{\includegraphics[scale=0.32]{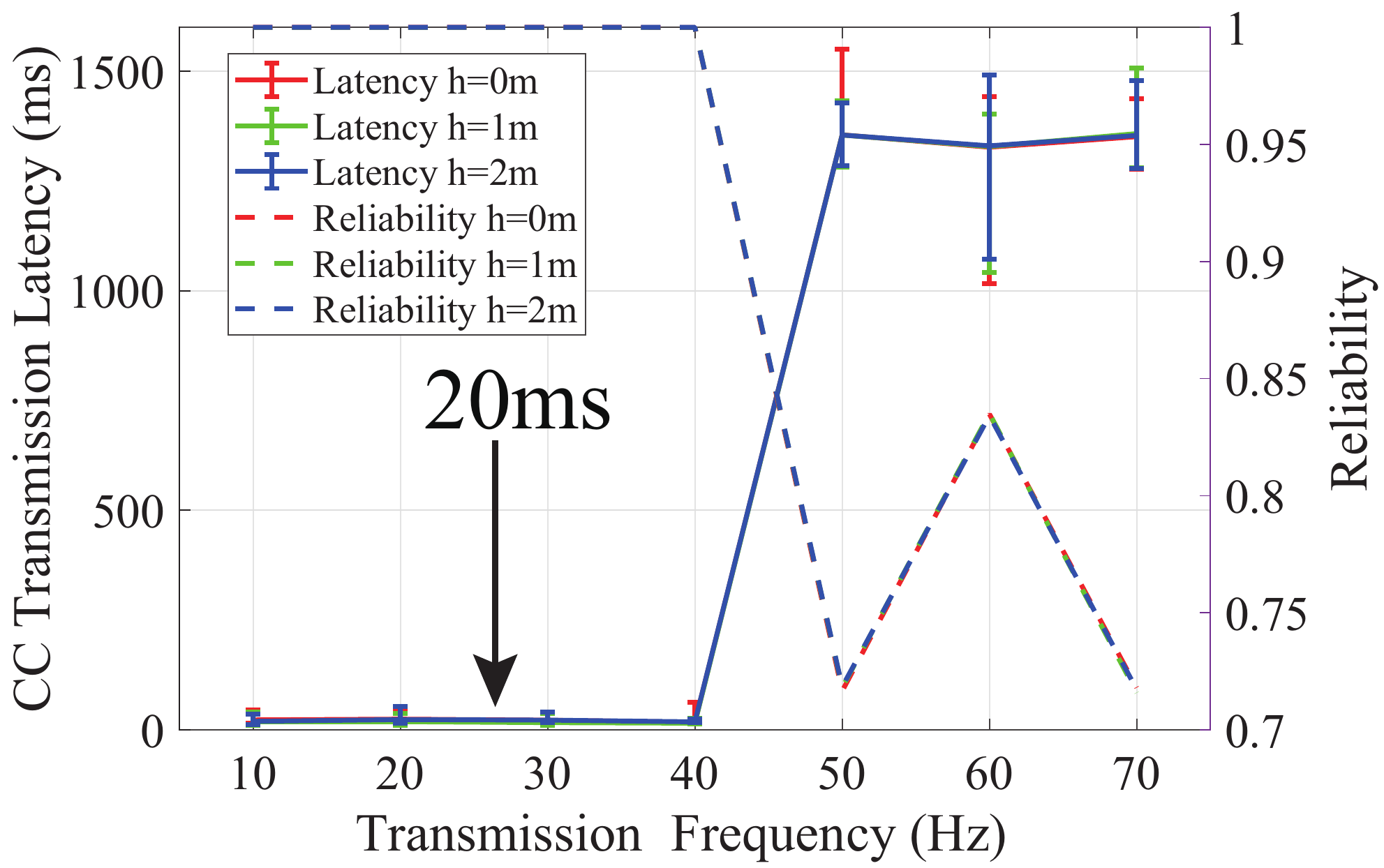}}
		\caption{E2E delay and reliability of CC transmission.}
		\label{Control latency}
	\end{figure}
	Fig.~\ref{Control latency} plots the E2E CC delay and reliability versus various transmission frequency for various flight heights, where the delay represents the elapsed time from the generation of CC packet at the EPC to the reception of CC packet at the UAV-UE. We set the control signal transmission frequency from 10Hz to 70Hz and calculates the E2E average delay, maximum delay, and minimum delay over $ 10^{4} $ control signal transmissions. We can see that the E2E CC delay and reliability are almost the same for flight heights 0m, 1m, and 2m. For a certain flight height, the E2E CC delay remains around 20ms during 10Hz to 40 Hz, and the corresponding reliability achieves 100\%. Surprisingly, the E2E CC delay increases rapidly to 1355ms after 40Hz, along with a significant decrease on the reliability. This is because the CC transmission frequency is higher than the CC receiving frequency as shown in Algorithm 2, leading to buffer overflow at the UAV-UE. We also see the tradeoff between the reliability and latency performances over low and high transmission frequency.

	\begin{figure}[!h]
		\centerline{\includegraphics[scale=0.29]{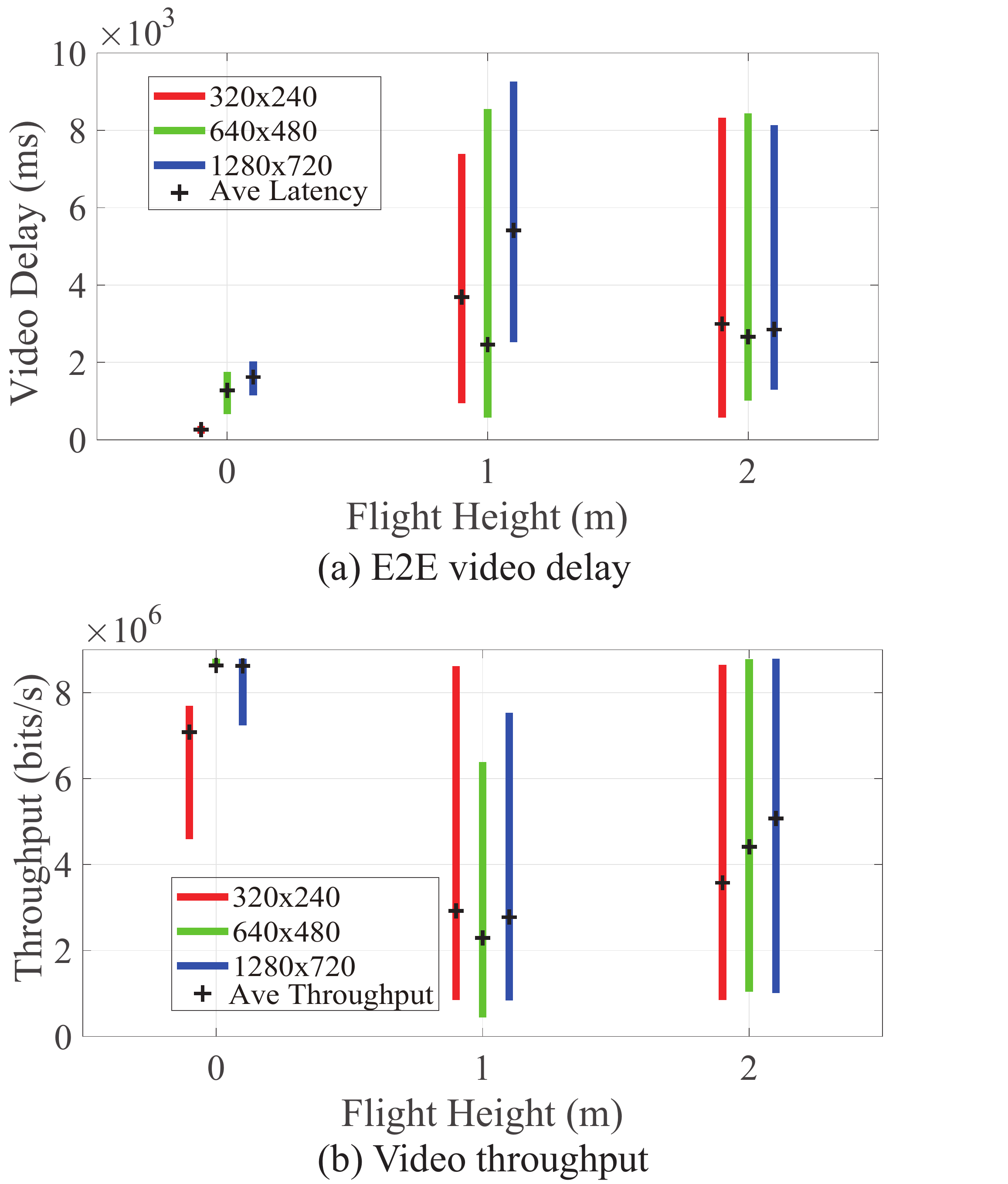}}
		\caption{E2E video transmission performance.}
		\label{Video Streaming Latency}
	\end{figure}
	Fig.~\ref{Video Streaming Latency} plots the E2E real-time video delay, and throughput versus the UAV-UE flight height for various video resolutions including 320x240, 640x480 and 1280x720. The E2E delay of video transmission represents the elapsed time from the capturing of image at the UAV-UE to the display of image at the EPC. For stationary UAV-UE ($h=$0m), the average E2E delay increases with higher resolution due to larger data size and longer encoding/decoding time consumption. We can see that the average throughput of 320x240 video is 7.08 Mb/s, and the average throughput of 640x480 and 1280x720 video achieve 8.6 Mb/s, which is limited by the uplink bandwidth in Table~\ref{Performance of the the established LTE network}. For flying UAV-UE ($h=$1m, 2m), the average E2E delay is more dependent on the network condition than video resolution due to UAV-UE's mobility. Since the  WebRTC is able to automatically adjust the FPS of the video based on the network condition, we can see that the average E2E delay of the 640x480 video transmission at $ h=$1m is lowest because of a lower FPS, which is consistent with its throughput performance. Overall, the UAV-UE flying at 2m has lower E2E video delay and higher throughput than that of UAV-UE flying at 1m because of a stronger LOS.

\begin{figure}[!htb]
	\centerline{\includegraphics[scale=0.42]{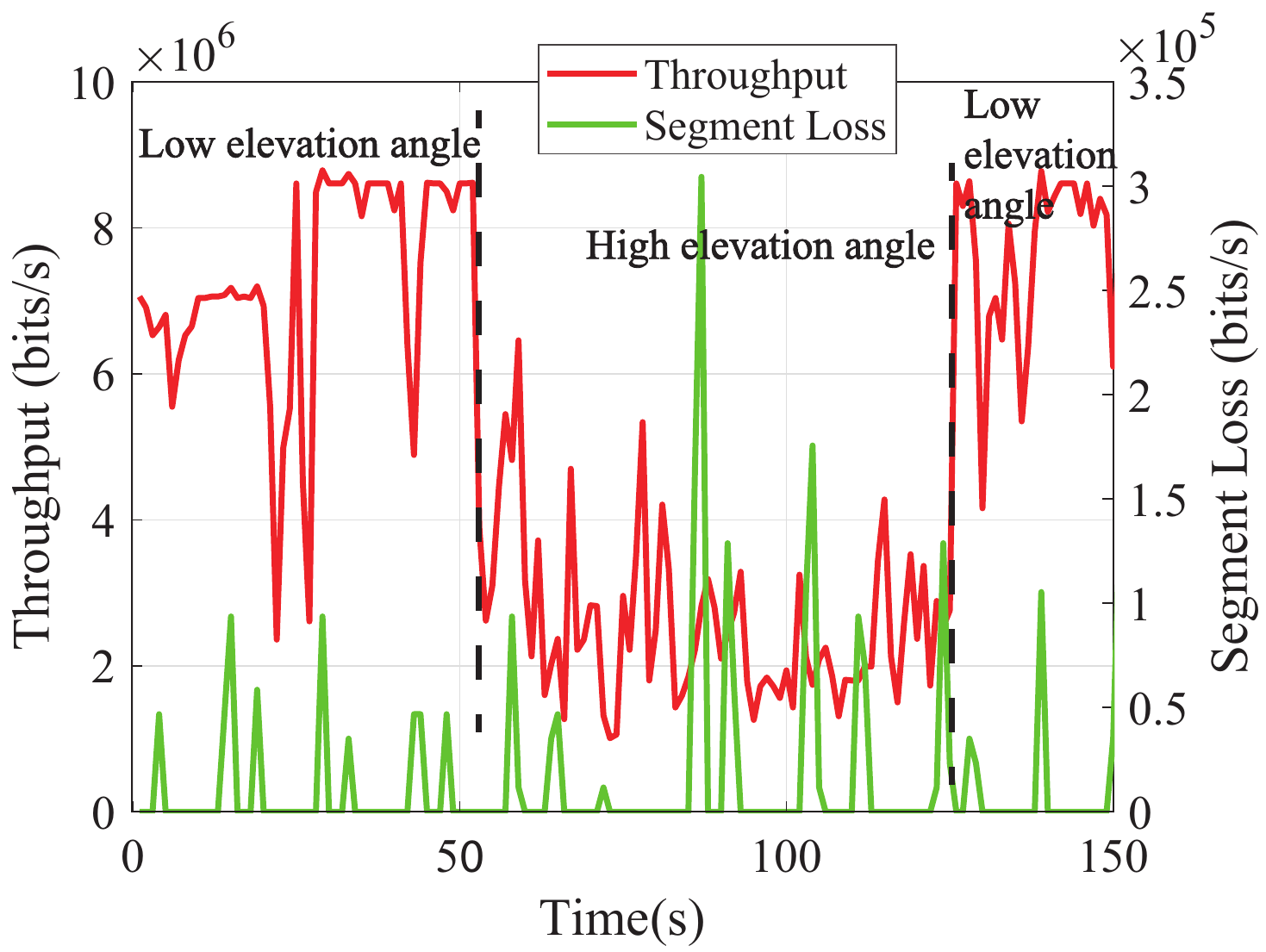}}
	\caption{E2E video transmission throughput and segment loss at different elevation point.}
	\label{coverage and error}
\end{figure}

Fig.~\ref{coverage and error} plots the E2E video transmission throughput and segment loss when the UAV flies at high elevation angle (i.e., close to the top of the BS) and low elevation angle (i.e., away from the top of the BS) positions, separately. The UAV-UE flies close to the BS first, and then flies away from the BS. The flight height and video resolution are $ \mathrm{2m} $ and $ 1280\times720 $, respectively. We can see that the throughput drops significant to $ 2\times10^{6}~\mathrm{bits/s} $ when the UAV-UE is at the high elevation angle position, and increases to $ 8.5\times10^{6}~\mathrm{bits/s} $ after returning to low elevation angle position. We can also observe the tradeoff between the throughput and segment loss, where the segment loss increases at high elevation angle and decreases at low elevation angle. This is because the tilted-down antenna of BS provides limited gain for UAV with high elevation angle, which is consistent with the analytical results in \cite{Mobilitysky}.




	\section{Conclusion}
	In this paper, we developed a cellular-connected UAV testbed to evaluate the throughput, E2E delay, and reliability of CC and real-time video streaming transmission. We first implemented the LTE network via OAI and USRP, and equipped the UAV with Raspberry Pi and LTE module to act as UAV-UE. We then established the downlink CC transmission and the uplink video transmission, and proposed corresponding schemes for the throughput, E2E delay, and reliability measurement. Our indoor experimental results have shown: 1) the buffer overflow limits the CC transmission frequency; 2) the FPS needs to be dynamically adapted to guarantee the continuity of video transmission service; and 3) the link outage problem caused by the BS antenna pattern remains an urgent issues to be solved.
	
	
	\bibliographystyle{IEEEtran} 
	
	\bibliography{testbed}
	
\end{document}